# Problem Solving in Sprego

Maria Csernoch, Piroska Biró
4028 Kassaiút. 26. Debrecen, Hungary
csernoch.maria@inf.unideb.hu, biro.piroska@inf.unideb.hu

**ABSTRACT**

*Sprego is a programming tool for novice and end-user programmers within graphical spreadsheet environments. The main idea of Sprego is to use as few general purpose functions as possible, and based on these functions we create multilevel formulas to solve real world programmable spreadsheet problems. Beyond providing the framework for the theoretic background and the tools which support Sprego, in order to demonstrate the power which lies within it, we present a converted authentic table and, based on this table, data retrieval tasks, their algorithms, and coding in full details.*

## 1    INTRODUCTION

When testing first year students of Informatics, we have found that the spreadsheet is not considered a programming environment. Students think that serious programmers have nothing to do with birotical (office application) software and documents. However, it is not only students but also researchers who consider the management of birotical documents a form of low-level routine knowledge [Bell and Newton, 2013], and a subject which is responsible for the failure to teach computers and informatics, and should be banished from school [Gove, 2012, 2014].This is definitely not the case. These claims prove that computational thinking [Wing, 2006] and deep approach computer problem solving [Csernoch and Biró, 2014, 2015a, 2015b, 2015c, 2015d] have not reached e-document handling and management.

Our research in testing the different problem solving approaches and methods [Biró et al., 2015a, 2015b; Csernoch and Biró, 2013, 2014, 2015a, 2015b] proved that document management, considered as a form of problem solving, does not differ from problem solving either in programming or in other sciences. The widespread and popular surface approach methods, disguised as user friendly methods, do not work, and they lead to the current high number of error prone documents, which waste human and computer resources and time [Panko, 2008; Tort et al., 2008; Van Deursen and Van Dijk, 2012]. On the other hand, the deep approach methods are proven to be efficient. These methods are concept- or CAAD-based (Computer Algorithmic And Debugging) [Csernoch and Biró, 2014, 2015b], and as such, the problem is in the focus, not the tools.

One of the consequences of the claims of the user friendly spreadsheet environments is that both students and end-users are lost in the enormous number of functions. On one hand, they are taught an extremely high number of functions. On the other hand, they do not understand the descriptions of the functions and their arguments in the wizards and help pages, because they do not have the vocabulary.

When completing the task "List the 15 spreadsheet functions which you think are the most important", first year students of informatics named 99 functions. To find the reasons for this incredibly high number of "important" functions, we checked the coursebooks written on spreadsheets, and 171 functions were found. Isn't this a frightening discovery? Further analysis of the spreadsheet coursebooks has led us to the conclusion that most of the authors are not able to distinguish between user manuals and coursebooks. These books do not provide meaningful examples, if they offer any at all;

they do nothing else but list the name and the argument list of the functions. Essentially, they repeat the wizards and the help pages with all their errors.

## 2 SPREGO

Sprego [Csernoch, 2014; Csernoch and Biró, 2015a, 2015d] is a deep approach problem solving programming method in the spreadsheet environment. It was proved around the early 90's [Booth,1992] that functional languages serve as a perfect first language, students being familiar with the concept of function from mathematics [Wakeling, 2007; Sestoft, 2011], and because of the simplicity of the language involved. The great advantage of these languages is that the focus is on the problem, not on the details of the coding [Csernoch and Balogh, 2010; Csernoch, 2012; Csernoch, 2014; Csernoch and Biró, 2015a, 2015d]. Sprego works on already existing tables, consequently, on data management and retrieval, and as such, focuses on the programming aspects of creating formulas. In general, Sprego is a programming tool within spreadsheets, supported by the following tools: Sprego functions, multilevel formulas, array formulas, debugging, and handling authentic sources.

### 2.1 Sprego functions

We use as few general purpose functions as possible (Table 1). The Sprego functions are classified based on the development made by the students. Sprego1 and Sprego2 consist of the dozen functions which are required. Sprego1 is for beginners, while the functions of Sprego2 are more demanding. Sprego3 is for intermediate and advanced students and end users. We have to emphasize here that all the three groups include general purpose functions.

Table 1. Sprego functions

| **Sprego1** | **Sprego2** | **Sprego3** |
|---|---|---|
| SUM() | MATCH() | SMALL() |
| AVERAGE() | INDEX() | LARGE() |
| MIN() | ISERROR() | AND() |
| MAX() | | OR() |
| LEFT() | | NOT() |
| RIGHT() | | ROW() |
| LEN() | | COLUMN() |
| SEARCH() | | OFFSET() |
| IF() | | SUBSTITUTE() |
| | | TRANSPOSE() |
| | | ROUND() |
| | | RAND() |
| | | INT() |

### 2.2 Multilevel formulas

Due to the low number of Sprego functions there is a need to create multilevel formulas. Multilevel formulas are required to solve real programming tasks.

### 2.3 Array formulas

Array formulas are those formulas whose output is an array, or which accept array(s) as argument(s), while the default arguments are single values. In the first case the output of the formula is an array– array result array formulas, ARAF –, while in the second case both array and single result – single result array formulas, SRAF –outputs are possible. With the use of array formulas, the copying of formulas and its consequences, which is one of the main sources of errors, is ruled out.

### 2.4 Debugging

The first three tools evoke a fourth one, debugging. The combination of the Sprego functions with the multilevel array formulas makes thorough debugging available in spreadsheets. Similar to high level programming languages, both manual and spreadsheet supported debugging can be carried out.

### 2.5 Authentic sources

By using authentic tables in the teaching of Sprego we would avoid the fiasco of other school programming languages. Most of the students find the programming languages designed for educational purposes useless. They are not able to transfer knowledge from the school programs to solve real word problems in other environments. With Sprego we can provide the students with real world problems; consequently, there is no gap between the student and the end user statuses.

In Sprego programming we follow the phases of the well accepted deep approach problem solving methods [Polya, 1954; Booth, 1992; Case&Gunstone, 2002; IEEE&ACM Report, 2013; Csernoch&Biró, 2013, 2014, 2015b]: (1) Based on the available data and being aware of the expected output we create a plan, (2) following this we carry out the coding, and the final step is (3) the discussion of the problem.

- Familiarity: Understanding the concept, seeing clearly what is required, seeing how the various items are connected, how the unknown is connected to the data, building the algorithm.

- Usage: Carrying out the plan, which, in a programming environment, is the coding of the algorithm.

- Assessment: Looking back at the complete solution, considering a concept from multiple viewpoints, justifying the selection of a particular approach, discussing, debugging it.

In the following section we provide a table downloaded from the LOL (League of Legends) message board [LOL, 2015] converted to a spreadsheet table. Based on these data, we present several tasks (Tasks 1–6) with their detailed solutions. Solving these problems we follow strictly the consecutive steps of deep approach problem solving: (1) understanding the problem, the concept, (2) building the algorithm, (3) coding, and (4]) debugging.

### 3    TASKS AND THEIR SPREGO SOLUTIONS

The converted LOL message board table consists of five columns and 1001 rows, where the first row holds the column titles (Figure 1 and Figure 2). The five columns consist of the following data.

- A: The title of the message.
- B: The name of the account and its server, where the message arrived from.
- C: The classification of the message (theme) and how long ago it was posted.
- D: The number of comments (NOF comments), with an accompanying string.
- E: The number of views (NOF views), with an accompanying string.

|      | A | B |
|------|---|---|
| 1    | Title | Account (server) |
| 2    | Arcade Sona's future (forums.euw.leagueoflegends.com) | ReisenII (EUW) |
| 4    | Season 5 changes have made League incredibly bland, dull, a | Maximum Kawaii (EUNE) |
| 10   | THE EU FORUM AWARDS | Riot Draggles (EUW) |
| 17   | [BUG] decreased framerate patch 5.4 | Papa Lovegood (EUW) |
| 18   | ( ͡° ͜ʖ ͡°) URGOT BUFFS ON PBE ( ͡° ͜ʖ ͡°) | DahakaGG (EUW) |
| 30   | Feedback to riot employees (youtube.com) | proapllegamer (EUNE) |
| 33   | veigar | BBS CursedSoul (EUW) |
| 34   | Give riven energy. | DarkSliceOfCake (EUW) |
| 35   | Try the champion before you buy them | CandyLandRemixed (EUW) |
| 54   | I always flame. What should i do? | filojistoNNN (EUW) |
| 55   | @ RIOT | trojanfighter (EUW) |
| 394  | Urgot Rework Idea | MB Ghost 2 Ghost (EUW) |
| 395  | New splash arts are WORSE! | GingarPowar (EUW) |
| 1001 | My account has been changed to Russia due to the most likel | NinjaJesus720 (EUW) |

Figure 1. Columns A–B of the converted LOL table

|      | C | D | E |
|------|---|---|---|
| 1    | Theme and time | NOF comments | NOF views |
| 2    | Champions & Gameplay about 18 hours ago | 14 new Comments | 680 Views |
| 4    | Champions & Gameplay about 4 hours ago | 14 new Comments | 149 Views |
| 10   | Forum Games & Contests about 24 hours ago | 32 new Comments | 1.1k Views |
| 17   | Suggestions & Bug Reports about 6 hours ago | 4 new Comments | 59 Views |
| 18   | Champions & Gameplay about 17 hours ago | 11 new Comments | 412 Views |
| 30   | Off-topic about 22 hours ago | 13 new Comments | 269 Views |
| 33   | Champions & Gameplay about 15 hours ago | 5 new Comments | 82 Views |
| 34   | Champions & Gameplay 2 days ago | 125 new Comments | 2.5k Views |
| 35   | Community Creations about 7 hours ago | 7 new Comments | 131 Views |
| 54   | Player Behaviour 44 minutes ago | 9 new Comments | 39 Views |
| 55   | Player Behaviour about an hour ago | 0 new Comments | 11 Views |
| 394  | Champions & Gameplay a day ago | 8 new Comments | 52 Views |
| 395  | Off-topic 3 days ago | 77 new Comments | 1.6k Views |
| 1001 | Help & Support 3 days ago | 1 new Comment | 147 Views |

Figure 2. Columns C–E of the converted LOL table

### 3.1 Array result array formulas

| Task 1 | Write out the name of the account in a separate column. |
|--------|--------------------------------------------------------|
| Solution 1 | S1–S3, Table 2. |

**Characteristics of Task 1**

- Accounts are on the left side of the strings.
- They are of different lengths.
- Accounts are followed by a Space and an opening parenthesis characters.

**Algorithm of Task 1**

- Deciding on the position of the ( character. Output: a number. S1
- Calculating the length of the account. Output: a number. S2
- Cutting out the account. Output: a string. S3

**Coding of Task 1**

S1. {=FIND("(",C2:C1001)}
S2. {=FIND("(",C2:C1001)-2}
S3. {=LEFT(C2:C1001,FIND("(",C2:C1001)-2)}

Table 2. The input and consecutive outputs of the solution of Task 1

| Account (server) | S1 | S2 | S3 |
| --- | --- | --- | --- |
| ReisenII (EUW) | 10 | 8 | ReisenII |
| Maximum Kawaii (EUNE) | 16 | 14 | Maximum Kawaii |
| Riot Draggles (EUW) | 15 | 13 | Riot Draggles |
| Papa Lovegood (EUW) | 15 | 13 | Papa Lovegood |
| DahakaGG (EUW) | 10 | 8 | DahakaGG |
| Proapllegamer (EUNE) | 15 | 13 | proapllegamer |
| BBS CursedSoul (EUW) | 16 | 14 | BBS CursedSoul |
| DarkSliceOfCake (EUW) | 17 | 15 | DarkSliceOfCake |
| CandyLandRemixed (EUW) | 18 | 16 | CandyLandRemixed |
| filojistoNNN (EUW) | 14 | 12 | filojistoNNN |
| trojanfighter (EUW) | 15 | 13 | trojanfighter |
| MB Ghost 2 Ghost (EUW) | 18 | 16 | MB Ghost 2 Ghost |
| GingarPowar (EUW) | 13 | 11 | GingarPowar |
| NinjaJesus720 (EUW) | 15 | 13 | NinjaJesus720 |

| Task 2 | Write out the number of comments without the text. |
| --- | --- |
| Solution 2 | Solution: S4–S7, Table 3 |

**Characteristics of Task 2**

- Numbers
- They are on the left side of the string
- They have different numbers of digits.
- Following the numbers there is a Space and an n character (the first character of *new comments*).

**Algorithm of Task 2**

- Finding the position of the n character. Output: a number.S4
- Calculating the length of the number. Output: a number.S5
- Cutting out the number from the original string. Output: a string.S6
- Converting the text into a number. Output: a number.S7

**Coding of Task 2**

S4.  {=FIND("new",E2:E1001)}
S5.  {=FIND("new",E2:E1001)-2}
S6.  {=LEFT(E2:E1001,FIND("new",E2:E1001)-2)}
S7.  {=LEFT(E2:E1001,FIND("new",E2:E1001)-2)*1}

Table 3. The input and consecutive outputs of the solution of Task 2

| NOF comments | S4 | S5 | S6 | S7 |
|---|---|---|---|---|
| 14 new Comments | 4 | 2 | 14 | 14 |
| 14 new Comments | 4 | 2 | 14 | 14 |
| 32 new Comments | 4 | 2 | 32 | 32 |
| 4 new Comments | 3 | 1 | 4 | 4 |
| 11 new Comments | 4 | 2 | 11 | 11 |
| 13 new Comments | 4 | 2 | 13 | 13 |
| 5 new Comments | 3 | 1 | 5 | 5 |
| 125 new Comments | 5 | 3 | 125 | 125 |
| 7 new Comments | 3 | 1 | 7 | 7 |
| 9 new Comments | 3 | 1 | 9 | 9 |
| 0 new Comments | 3 | 1 | 0 | 0 |
| 8 new Comments | 3 | 1 | 8 | 8 |
| 77 new Comments | 4 | 2 | 77 | 77 |
| 1 new Comment | 3 | 1 | 1 | 1 |

| Task 3 | Write out the name of the server. |
|---|---|
| Solution 3 | Solution: S8–S12, **Table 4** |

**Characteristics of Task 3**

- Servers are on the right side of the original string.
- They are in a pair of parentheses.
- They are texts.
- They are of different lengths.

**Algorithm of Task 3**

- Calculating the length of the original string, with account and server together. Output: a number.S8

- Calculating the difference between the length of the text and the position of the opening parenthesis. Output: a number.S9

- Cutting out the name of the server and the closing parenthesis from the right side of the string. Output: a string.S10

- Calculating the length of the short string. Output: a number.S11

- Cutting out the name of the server. Output: a string.S12

**Coding of Task 3**

S8. {=LEN(C2:C1001)}
S9. {=LEN(C2:C1001)-FIND("(",C2:C1001)}
S10. {=RIGHT(C2:C1001,LEN(C2:C1001)-FIND("(",C2:C1001))}
S11. {=LEN(RIGHT(C2:C1001,LEN(C2:C1001)-FIND("(",C2:C1001)))}
S12. {=LEFT(RIGHT(C2:C1001,LEN(C2:C1001)-FIND("(",C2:C1001)),
      LEN(RIGHT(C2:C1001,LEN(C2:C1001)-FIND("(",C2:C1001)))-1)}

Table 4. The input and consecutive outputs of the solution of Task 3

| Account (server) | S8 | S9 | S10 | S11 | S12 |
|---|---|---|---|---|---|
| ReisenII (EUW) | 14 | 4 | EUW) | 4 | EUW |
| Maximum Kawaii (EUNE) | 21 | 5 | EUNE) | 5 | EUNE |
| Riot Draggles (EUW) | 19 | 4 | EUW) | 4 | EUW |
| Papa Lovegood (EUW) | 19 | 4 | EUW) | 4 | EUW |
| DahakaGG (EUW) | 14 | 4 | EUW) | 4 | EUW |
| proapllegamer (EUNE) | 20 | 5 | EUNE) | 5 | EUNE |
| BBS CursedSoul (EUW) | 20 | 4 | EUW) | 4 | EUW |
| DarkSliceOfCake (EUW) | 21 | 4 | EUW) | 4 | EUW |
| CandyLandRemixed (EUW) | 22 | 4 | EUW) | 4 | EUW |
| filojistoNNN (EUW) | 18 | 4 | EUW) | 4 | EUW |
| trojanfighter (EUW) | 19 | 4 | EUW) | 4 | EUW |
| MB Ghost 2 Ghost (EUW) | 22 | 4 | EUW) | 4 | EUW |
| GingarPowar (EUW) | 17 | 4 | EUW) | 4 | EUW |
| NinjaJesus720 (EUW) | 19 | 4 | EUW) | 4 | EUW |

| Task 4 | Write out the number of views without the text. |
|---|---|
| Solution 4 | Solution: S13–S18, Table **5**, S1–S5, Table 6 |

**Characteristics of Task 4**

- Numbers followed by the string Views.

- Two different kinds of numbers: (1) whole numbers, (2) real numbers in thousands, marked by k.

- Numbers are on the left side of the original string.

**Algorithm of Task 4 (1st section)**

- Finding the position of the V character (the first character of Views). Output: a number. S13

- Calculating the length of the number. Output: a number. S14

- Cutting out the number from the original string. Output: a string. S15

- Converting the string to a number. Two different outputs: (1) a number, (2) an error message. The formula works on the whole numbers. S16

- Shortening the S15 string by one character. The original purpose of the removal of the k character from the end of the thousand numbers. Two different outputs: (1) whole numbers one digit shorter, (2) thousand numbers, as a real number [without the k character]. S17

- Multiplying the numbers by 1000. Two different outputs: whole numbers 100 times greater than the original value, (2) kilos are converted to a rounded number. S18

**Coding of Task 4 [1st section]**

S13. {=FIND("V",F2:F1001)}
S14. {=FIND("V",F2:F1001)-2}
S15. {=LEFT(F2:F1001,FIND("V",F2:F1001)-2)}
S16. {=LEFT(F2:F1001,FIND("V",F2:F1001)-2)*1}
S17. {=LEFT(F2:F1001,FIND("V",F2:F1001)-3)*1}
S18. {=LEFT(F2:F1001,FIND("V",F2:F1001)-3)*1000}

Table 5. The input and consecutive outputs (S13–S18) of the solution of Task 4

| NOF Views | S13 | S14 | S15 | S16 | S17 | S18 |
|---|---|---|---|---|---|---|
| 680 Views | 5 | 3 | 680 | 680 | 68 | 68000 |
| 149 Views | 5 | 3 | 149 | 149 | 14 | 14000 |
| 1.1k Views | 6 | 4 | 1.1k | #VALUE! | 1.1 | 1100 |
| 59 Views | 4 | 2 | 59 | 59 | 5 | 5000 |
| 412 Views | 5 | 3 | 412 | 412 | 41 | 41000 |
| 269 Views | 5 | 3 | 269 | 269 | 26 | 26000 |
| 82 Views | 4 | 2 | 82 | 82 | 8 | 8000 |
| 2.5k Views | 6 | 4 | 2.5k | #VALUE! | 2.5 | 2500 |
| 131 Views | 5 | 3 | 131 | 131 | 13 | 13000 |
| 39 Views | 4 | 2 | 39 | 39 | 3 | 3000 |
| 11 Views | 4 | 2 | 11 | 11 | 1 | 1000 |
| 52 Views | 4 | 2 | 52 | 52 | 5 | 5000 |
| 1.6k Views | 6 | 4 | 1.6k | #VALUE! | 1.6 | 1600 |
| 147 Views | 5 | 3 | 147 | 147 | 14 | 14000 |

**Algorithm of Task 4 [2nd section]**

- Searching for the k character in the string. Two different outputs: (1) error message with whole numbers, (2) a number, the position of the k character with real numbers. S1

- Checking the output of the formula which searches for the k character. The question is whether the k is found or not. Output: (1) TRUE, with the whole numbers, no k was found, the function searching for k returned with an error, (2) FALSE, with the real numbers, the function searching for k returned with a correct output, there is no error. S2

- Setting a yes/no question based on the results of the check for an error. Output: zeros, because the question is set, but none of the answers. S3

- Setting the output, if the answer is yes to the question. The answer is copied from S16. Output: (1) whole numbers and (2) 0s. S4

- Setting the output, if the answer is no to the question. The answer is copied from S17. Output: (1) whole numbers and (2) numbers rounded to hundreds. S5

Table 6. The second section (S1–S5) of the solution of Task 4

| S1 | S2 | S3 | S4 | S5 |
|---|---|---|---|---|
| #VALUE! | TRUE | 0 | 680 | 680 |
| #VALUE! | TRUE | 0 | 149 | 149 |
| 4 | FALSE | 0 | 0 | 1100 |
| #VALUE! | TRUE | 0 | 59 | 59 |
| #VALUE! | TRUE | 0 | 412 | 412 |
| #VALUE! | TRUE | 0 | 269 | 269 |
| #VALUE! | TRUE | 0 | 82 | 82 |
| 4 | FALSE | 0 | 0 | 2500 |
| #VALUE! | TRUE | 0 | 131 | 131 |
| #VALUE! | TRUE | 0 | 39 | 39 |
| #VALUE! | TRUE | 0 | 11 | 11 |
| #VALUE! | TRUE | 0 | 52 | 52 |
| 4 | FALSE | 0 | 0 | 1600 |
| #VALUE! | TRUE | 0 | 147 | 147 |

**Coding of Task 4 [2nd section]**

S1. {=FIND("k",F2:F1001)}
S2. {=ISERROR(FIND("k",F2:F1001))}
S3. {=IF(ISERROR(FIND("k",F2:F1001)),,)}
S4. {=IF(ISERROR(FIND("k",F2:F1001)),LEFT(F2:F1001,FIND("V",F2:F1001)-2)*1,)}
S5. {=IF(ISERROR(FIND("k",F2:F1001)),LEFT(F2:F1001,FIND("V",F2:F1001)-2)*1,LEFT(F2:F1001,FIND("V",F2:F1001)-3)*1000)}

### 3.2 Conditional single result array formulas [CSRAF]

One of the most error prone classes of functions is the conditional built-in functions, found in different categories in spreadsheets. For a more convenient reference, we have created the *IF?() expression. The class of *IF?() functions holds all the conditional spreadsheet functions. To handle conditions the database function would also be appropriate, but they are even more burdensome than the *IF?() functions.

CSRAF scan substitute either the *IF?() or the database functions. Due to the limitations and the inconsistencies of these built-in functions [Csernoch, 2014; Csernoch and Biró; 2015a] CSRAFs offer a much wider variety of options, consequently they can be used to solve problems which we have never considered within the traditional spreadsheet framework.

| Task 5 | Type a number in H1003. How many messages have received more than H1003 views? (We use the output vector of Task 4, stored in column I.) |
|---|---|
| Solution 5 | Solution: S6–S8, Table 7 |

**Characteristics of Task 5**

- We have to separate those messages which have received more than H1003 views from those which have received fewer.

- We have to mark those messages which have received more than H1003 views. Each time a message is received we draw a little stick on a piece of paper. On a computer the easiest method is to store 1 for each match. If there is no match we leave the message unnoticed.

**Algorithm of Task 5**

- Asking 1000 questions whether the "NOF views" is greater than H1003, or not. Output: a vector of 1000 components of TRUE's and FALSE's, whose first component is displayed in the cell.S6

- Deciding on the output if the answer is yes: 1. Deciding on the output if the answer is no: default FALSE. Output: a vector of 1s and FALSE's, whose first component is displayed in the cell.S7

- Summing the components of the vector. Output: a whole number, the number of 1s stored in the vector.S8

**Coding of Task 5**

S6. {=I2:I1001>H1003}
S7. {=IF(I2:I1001>H1003,1)}
S8. {=SUM(IF(I2:I1001>H1003,1))}

Table 7. The first component of the output vector with two possible inputs (H1003) in Task 5

| NOF Views | H1003 | S6 | S7 | S8 |
|---|---|---|---|---|
| 680 Views | 500 | TRUE | 1 | 47 |
| 680 Views | 1600 | FALSE | FALSE | 11 |

| Task 6 | Type a server in G1004. Give the average and maximum of comments to messages arriving from the G1004 server. (We use the output vector of Task 3,servers stored in column G, and the output vector of Task 2, NOF comments stored in column H.) |
|---|---|
| Solution 6 | Solution: S9–S11/S12, Table 8 |

To solve Task 6 we use exactly the same algorithm as we did in Task 5.

**Characteristics of Task 6**

- We need to store only those "NOF comments" which arrived to messages coming from the server given in G1004.

- The average/maximum of these numbers should be calculated.

**Algorithm of Task 6**

- Asking the question. Output: a vector of TRUE's and FALSE's, whose first component is displayed in the cell.S9

- Deciding on the output of questions. Output: a vector of whole numbers [NOF comments] and FALSE's, whose first component is displayed in the cell.S10

- Calculating the average/maximum of the components of the vector. Output: a real/whole number, the average/maximum of "NOF comments" stored in the vector.S11/S12

**Coding of Task 6**

S9. {=G2:G1001=G1004}
S10. {=IF(G2:G1001=G1004,H2:H1001)}
S11. {=AVERAGE(IF(G2:G1001=G1004,H2:H1001))}
S12. {=MAX(IF(G2:G1001=G1004,H2:H1001))}

Table 8: The first component of the output vector with two possible inputs (G1004) in Task 6

| Account (server) | G1004 | S9 | S10 | S11 | S12 |
|---|---|---|---|---|---|
| ReisenII (EUW) | EUW | TRUE | 14 | 5.17 | 130 |
| ReisenII (EUW) | EUWE | FALSE | FALSE | 4.59 | 76 |

## 4  CONCLUSION

Sprego is a programming method within the spreadsheet framework. The main characteristics of Sprego are the use of general purpose functions and the building of multilevel formulas based on these functions. Our testing and analyses [Biró and Csernoch, 2013a, 2013b, 2014a, 2014b; Csernoch and Biró, 2013, 2014] have proved that Sprego can be used effectively both as a first programming language and as the language of end-user programmers. Due to the simplicity of the language and the few functions in use, with Sprego we can build up knowledge in long term memory, which would lead to less error prone documents. Beyond the reliability and stability of the documents, Sprego formulas are version and application independent. The documents can be freely opened both in MS Excel and OpenOffice/LibreOfficeCalc, without the need to check the versions of these programs.

In Sprego programming we handle problems in a way that is well accepted in traditional sciences and traditional programming languages, where the deep approach methods have proved effective. Sprego fulfills all the requirements of the deep approach metacognitive problem solving methods. The tasks and their solutions presented in this paper clearly demonstrate that programming in non-traditional environments would be as effective as in traditional programming languages.

## 5  ACKNOWLEDGEMENT

The research was supported partly by the TÁMOP-4.1.2.B.2-13/1-2013-0009, SZAKTÁRNET projects. The project has been supported by the European Union, co-financed by the European Social Fund. The research was supported partly by the Hungarian Scientific Research Fund under Grant No. OTKA K-105262.